\input harvmac.tex
%

\figno=0
\def\fig#1#2#3{
\par\begingroup\parindent=0pt\leftskip=1cm\rightskip=1cm\parindent=0pt
\baselineskip=11pt \global\advance\figno by 1 \midinsert
\epsfxsize=#3 \centerline{\epsfbox{#2}} \vskip 12pt {\bf Fig.
\the\figno:} #1\par
\endinsert\endgroup\par
}
\def\figlabel#1{\xdef#1{\the\figno}}
\def\encadremath#1{\vbox{\hrule\hbox{\vrule\kern8pt\vbox{\kern8pt
\hbox{$\displaystyle #1$}\kern8pt} \kern8pt\vrule}\hrule}}

\overfullrule=0pt

%
%

\font\zfont = cmss10 
 
\def\bigone{\hbox{1\kern -.23em {\rm l}}}
\def\ZZ{\hbox{\zfont Z\kern-.4emZ}}

\def\A{{\cal B}}
\def\a{\alpha}

\def\g{\gamma}
\def\d{\delta}
\def\e{\epsilon}

\def\th{\theta}

\def\k{\kappa}
\def\l{\lambda}

\def\p{\phi}

\def\c{\chi}

\def\G{\Gamma}

\def\O{\Omega}
\def\o{\over}

\def\w{\wedge}

\def\bz{{\bar z}}
\font\cmss=cmss10 \font\cmsss=cmss10 at 7pt
\def\rlx{\relax\leavevmode}

\def\RR{\relax{\rm I\kern-.18em R}}
\def\ZZ{\rlx\leavevmode\ifmmode\mathchoice{\hbox{\cmss Z\kern-.4em Z}}
 {\hbox{\cmss Z\kern-.4em Z}}{\lower.9pt\hbox{\cmsss Z\kern-.36em Z}}
 {\lower1.2pt\hbox{\cmsss Z\kern-.36em Z}}\else{\cmss Z\kern-.4em
 Z}\fi}

\def\ap{\alpha'}

\def\frac#1#2{{#1 \over #2}}
\def\pa{\partial}

\def\rd{{ d}}

\def\bpa{{\bar \partial}}

\def\ba{{\bar a}}
\def\bb{{\bar b}}

\def\tJ{{\tilde J}}
\def\hR{{\hat R}}
\def\om{{\omega}}
\def\bom{{\bar \omega}}
\def\btom{{\bar{\tilde \omega}}}
\def\tom{{\tilde \omega}}
\def\tL{{\tilde L}}
\def\hom{{\hat \omega}}
\def\CA{{\cal A}}

\def\CF{{\cal F}}

\def\CS{{\cal S}}

\def\na{\nabla}
\def\bth{{\bar \theta}}

\def\bO{{\bar \Omega}}

\def\IZ{\relax\ifmmode\mathchoice
{\hbox{\cmss Z\kern-.4em Z}}{\hbox{\cmss Z\kern-.4em Z}}
{\lower.9pt\hbox{\cmsss Z\kern-.4em Z}} {\lower1.2pt\hbox{\cmsss
Z\kern-.4em Z}}\else{\cmss Z\kern-.4em Z}\fi}


\lref\Strom{ A.~Strominger, ``Superstrings with torsion,'' Nucl.\
Phys.\ B {\bf 274}, 253 (1986). } \lref\Bars{ I.~Bars,
``Compactification of superstrings and torsion,'' Phys.\ Rev.\ D
{\bf 33}, 383 (1986);
I.~Bars, D.~Nemeschansky and S.~Yankielowicz, ``Compactified
superstrings and torsion,'' Nucl.\ Phys.\ B {\bf 278}, 632 (1986).
} \lref\Hull{ C.~M.~Hull, ``Anomalies, ambiguities and
superstrings,'' Phys.\ Lett.\ B {\bf 167}, 51 (1986).
} \lref\Hulla{ C.~M.~Hull, ``Sigma model beta functions and string
compactifications,'' Nucl.\ Phys.\ B {\bf 267}, 266 (1986);
C.~M.~Hull and E.~Witten, ``Supersymmetric sigma models and the
heterotic string,'' Phys.\ Lett.\ B {\bf 160}, 398 (1985);
}

\lref\Hullb{C.~M.~Hull, ``Compactifications of the heterotic
superstring,'' Phys.\ Lett.\ B {\bf 178}, 357 (1986).
} \lref\Sen{ A.~Sen, ``(2, 0) Supersymmetry And Space-Time
Supersymmetry In The Heterotic String Theory,'' Nucl.\ Phys.\ B
{\bf 278}, 289 (1986).
}

\lref\FuYau{ J.-X.~Fu and S.-T.~Yau, ``The theory of superstring
with flux on non-K\"ahler manifolds and the complex Monge-Amp\'ere
equation,'' [arXiv:hep-th/0604063].
}

\lref\LiYau{J.~Li and S.-T.~Yau, ``Hermitian-Yang-Mills connection
on non-K\"ahler manifolds,'' in {\it Mathematical aspects of
string theory}, World Scientific Publ., S.-T.~Yau, editor; London
560 (1987).}

\lref\Lubke{M.~L\"ubke and A.~Teleman, {\it The Kobayashi-Hitchin
correspondence}, World Scientific, River Edge, NJ (1995).}

\lref\Yau{S.-T.~Yau, ``On the Ricci curvature of a compact
K\"ahler manifold and the complex Monge-Amp\'ere equations. I,''
Comm.\ Pure\ Appl.\ Math.\ {\bf 31}, 339 (1978).}

\lref\UYau{K.~Uhlenbeck and S.-T.~Yau, ``On the existence of
Hermitian-Yang-Mills connections in stable vector bundles,''
Comm.\ Pure\ Appl.\ Math.\ {\bf 39}, S257 (1986).}

\lref\Lang{S.~Lang, {\it Real and functional analysis}, 3rd ed.,
Springer-Verlag, New York (1983). }

\lref\JostA{J.~Jost, {\it Partial differential equations},
Springer-Verlag, New York (2002).}

\lref\JostB{J.~Jost, {\it Postmodern analysis}, Springer-Verlag,
New York (1998).}

\lref\DasRS{K.~Dasgupta, G.~Rajesh and S.~Sethi, ``M theory,
orientifolds and $G$-flux,'' JHEP {\bf 9908}, 023 (1999)
[arXiv:hep-th/9908088].
}

\lref\BD{K.~Becker and K.~Dasgupta, ``Heterotic strings with
torsion,'' JHEP {\bf 0211}, 006 (2002) [arXiv:hep-th/0209077].
} \lref\BBDG{ K.~Becker, M.~Becker, K.~Dasgupta and P.~S.~Green,
``Compactifications of heterotic theory on non-K\"ahler complex
manifolds: I,'' JHEP {\bf 0304}, 007 (2003)
[arXiv:hep-th/0301161].
} \lref\BBDGS{ K.~Becker, M.~Becker, P.~S.~Green, K.~Dasgupta and
E.~Sharpe, ``Compactifications of heterotic strings on
non-K\"ahler complex manifolds: II,'' Nucl.\ Phys.\ B {\bf 678},
19 (2004) [arXiv:hep-th/0310058].
}

\lref\BBDGS{ K.~Becker, M.~Becker, P.~S.~Green, K.~Dasgupta and
E.~Sharpe, ``Compactifications of heterotic strings on
non-K\"ahler complex manifolds: II,'' Nucl.\ Phys.\ B {\bf 678},
19 (2004) [arXiv:hep-th/0310058].
}

\lref\DasRS{ K.~Dasgupta, G.~Rajesh and S.~Sethi, ``M theory,
orientifolds and $G$-flux,'' JHEP {\bf 9908}, 023 (1999)
[arXiv:hep-th/9908088].
}

\lref\BT{ K.~Becker and L.-S.~Tseng, ``Heterotic flux
compactifications and their moduli,'' Nucl.\ Phys.\ B {\bf 741},
162 (2006) [arXiv:hep-th/0509131].
}

\lref\GoldP{ E.~Goldstein and S.~Prokushkin, ``Geometric model for
complex non-K\"ahler manifolds with SU(3) structure,'' Commun.\
Math.\ Phys.\  {\bf 251}, 65 (2004) [arXiv:hep-th/0212307].
}

\lref\Verb{ M.~Verbitsky, ``Stable bundles on positive principal
elliptic fibrations,'' Math.\ Res.\ Lett.\ {\bf 12} 251 (2005)
[arXiv:math.AG/0403430]. }

\lref\Gaud{ P.~Gauduchon, ``La $1$-forme de torsion d'une
vari\'et\'e hermitienne compacte,'' [Torsion $1$-forms of compact
Hermitian manifolds] Math.\ Ann.\ {\bf 267} no. 4, 495 (1984).}

\lref\deWSH{ B.~de Wit, D.~J.~Smit and N.~D.~Hari Dass, ``Residual
supersymmetry of compactified D = 10 supergravity,'' Nucl.\ Phys.\
B {\bf 283}, 165 (1987).
}

\lref\GilT{ D.~Gilbarg and N.~S.~Trudinger, {\it Elliptic partial
differential equations of second order}, Springer-Verlag, New York
(1983).}

\lref\Mukai{S.~Mukai, ``Moduli of vector bundles on K3 surfaces,
and symplectic manifolds,'' Sugaku Expositions, {\bf 1}, 139
(1988).}

\lref\Dijk{R.~Dijkgraaf, ``Instanton strings and hyperKaehler
geometry,'' Nucl.\ Phys.\ B {\bf 543}, 545 (1999)
[arXiv:hep-th/9810210].
}

\lref\Yosh{ K.~Yoshioka, ``Irreducibility of moduli spaces of
vector bundles on $K3$ surfaces,'' [arXiv:math.AG/9907001].}

\lref\JLi{J.~Li, private communications.}

\lref\MaldaN{ J.~M.~Maldacena and C.~Nunez, ``Supergravity
description of field theories on curved manifolds and a no go
theorem,'' Int.\ J.\ Mod.\ Phys.\ A {\bf 16}, 822 (2001)
[arXiv:hep-th/0007018].
}

\lref\Lust{ G.~L.~Cardoso, G.~Curio, G.~Dall'Agata, D.~Lust,
P.~Manousselis and G.~Zoupanos, ``Non-Kaehler string backgrounds
and their five torsion classes,'' Nucl.\ Phys.\ B {\bf 652}, 5
(2003) [arXiv:hep-th/0211118].
} 
\lref\GauMW{ J.~P.~Gauntlett, D.~Martelli and D.~Waldram,
``Superstrings with intrinsic torsion,'' Phys.\ Rev.\ D {\bf 69},
086002 (2004) [arXiv:hep-th/0302158].
}

\lref\BeckerGJ{
  K.~Becker and M.~Becker,
  ``M-Theory on Eight-Manifolds,''
  Nucl.\ Phys.\ B {\bf 477}, 155 (1996)
  [arXiv:hep-th/9605053].
}

\lref\GreeneYA{
  B.~R.~Greene, A.~D.~Shapere, C.~Vafa and S.~T.~Yau,
  ``Stringy Cosmic Strings And Noncompact Calabi-Yau Manifolds,''
  Nucl.\ Phys.\ B {\bf 337}, 1 (1990).
}

\lref\StromingerIT{
  A.~Strominger, S.~T.~Yau and E.~Zaslow,
  ``Mirror symmetry is T-duality,''
  Nucl.\ Phys.\ B {\bf 479}, 243 (1996)
  [arXiv:hep-th/9606040].
}

\lref\AspinwallAD{
  P.~S.~Aspinwall and R.~Kallosh,
  ``Fixing all moduli for M-theory on K3 x K3,''
  JHEP {\bf 0510}, 001 (2005)
  [arXiv:hep-th/0506014].
}

\lref\grha{P.~Griffiths and J.~Harris, {\it Principles of
algebraic geometry}, Wiley-Interscience,  New York (1994).}

\lref\michel{M.~L.~Michelsohn, ``On the existence of special
metrics in complex geometry,'' Acta\ Math.\ {\bf 149}, 261
(1982).}

\lref\Bime{
  A.~Bilal and S.~Metzger,
  ``Anomaly cancellation in M-theory: A critical review,''
  Nucl.\ Phys.\ B {\bf 675}, 416 (2003)
  [arXiv:hep-th/0307152].}

\lref\schwarz{M.~B.~Green and J.~H.~Schwarz, ``Anomaly
cancellation in supersymmetric D=10 gauge theory and superstring
theory,'' Phys.\ Lett.\ B {\bf 149}, 117 (1984).
}

\lref\liyau{J.~Li and S.~T.~Yau, ``The existence of supersymmetric
string theory with torsion,'' J.\ Differential\ Geom.\  {\bf 70},
143 (2005) [arXiv:hep-th/0411136].
}

\lref\ovrut{B.~A.~Ovrut, ``Lectures on heterotic M-theory,'' in
{\it Strings, branes and extra dimensions: TASI 2001:
Proceedings}, ed. S.~S.~Gubser and J.~D.~Lykken, World Scientific,
New York, 359 (2004) [arXiv:hep-th/0201032].
}

\lref\dlm{M.~J.~Duff, J.~T.~Liu and R.~Minasian,
``Eleven-dimensional origin of string / string duality: A one-loop
test,'' Nucl.\ Phys.\ B {\bf 452}, 261 (1995)
[arXiv:hep-th/9506126].
}

\lref\ESharpe{ E.~Sharpe, ``Lectures on D-branes and sheaves,''
[arXiv:hep-th/0307245].
}

\lref\IPapad{
S.~Ivanov and G.~Papadopoulos,
``Vanishing theorems and string backgrounds,''
Class.\ Quant.\ Grav.\  {\bf 18}, 1089 (2001)
[arXiv:math.dg/0010038].
}

\lref\disgr{J.~Distler and B.~R.~Greene, ``Aspects Of (2,0) String
Compactifications,'' Nucl.\ Phys.\ B {\bf 304}, 1 (1988).}

\lref\gvw{
  S.~Gukov, C.~Vafa and E.~Witten,
  ``CFT's from Calabi-Yau four-folds,''
  Nucl.\ Phys.\ B {\bf 584}, 69 (2000)
  [Erratum-ibid.\ B {\bf 608}, 477 (2001)]
  [arXiv:hep-th/9906070].}

\lref\gkp{S.~B.~Giddings, S.~Kachru and J.~Polchinski,
  ``Hierarchies from fluxes in string compactifications,''
  Phys.\ Rev.\ D {\bf 66}, 106006 (2002)
  [arXiv:hep-th/0105097].}


\Title{ {\vbox{ \rightline{\hbox{hep-th/0604137}}
}}} {\vbox{ \hbox{\centerline{Anomaly Cancellation and Smooth
Non-K\"ahler}}\hbox{ } \hbox{\centerline{Solutions in Heterotic String
Theory}} }}
\bigskip
\centerline{Katrin Becker$^{1,2}$, Melanie Becker$^{1,2}$,
Ji-Xiang Fu$^{3,4}$, Li-Sheng Tseng$^{5,6}$ and Shing-Tung Yau$^6$}
\smallskip
\bigskip
\centerline{$^1$ \it George P. and Cynthia W. Mitchell Institute
for Fundamental Physics} \centerline{\it Texas A \& M University,
College Station, TX 77843, USA}
\smallskip
\centerline{$^2$\it Radcliffe Institute, Harvard University,
Cambridge, MA 02138, USA}
\smallskip
\centerline{$^3$\it School of Mathematical Sciences, Fudan
University, Shanghai, 200433, P.R. China}
\smallskip
\centerline{$^4$\it Key Laboratory of Mathematics for Nonlinear
Sciences} \centerline{\it Fudan University, Ministry of Education, P.R. China}
\smallskip
\centerline{$^5$\it Department of Physics, University of Utah,
Salt Lake City, UT 84112, USA}
\smallskip
\centerline{$^6$\it Department of Mathematics, Harvard University,
Cambridge, MA 02138, USA}

\bigskip
\bigskip

\bigskip

\centerline{\bf Abstract}

\bigskip

We show that six-dimensional backgrounds that are $T^2$ bundle
over a Calabi--Yau two-fold base are consistent smooth solutions
of heterotic flux compactifications. We emphasize the importance
of the anomaly cancellation condition which can only be satisfied
if the base is $K3$ while a $T^4$ base is excluded. The conditions
imposed by anomaly cancellation for the $T^2$ bundle structure, the
dilaton field, and the holomorphic stable bundles are analyzed and
the solutions determined. Applying duality, we check the
consistency of the anomaly cancellation constraints with those for
flux backgrounds of M-theory on eight-manifolds.

\bigskip
\baselineskip 18pt
\bigskip
\noindent

\Date{April, 2006}

\newsec{Introduction}

Since their discovery, almost ten years ago, tractable flux
compactifications in string theory have become a very active area
of research. The reasons for this are numerous but they share the
common feature of putting the connection between string theory and
realistic models of particle phenomenology into a new focus. Some
of the most vexing problems in high energy physics, like the
cosmological constant problem, moduli stabilization or the
hierarchy problem, have found a natural description within string
theory once fluxes are taken into account.

However, besides intense work on flux backgrounds in string
theory, the properties of the spacetime geometry, is largely an
uncharted territory. The conditions imposed by supersymmetry have
been understood in detail, however, less is known about the
background geometries, especially for the interesting case of the
heterotic string with fluxes. Generically, the presence of
$H$-flux in compactifications of the heterotic string is required
due to the presence of $\alpha'$ corrections in the Bianchi
identity which are needed for anomaly cancellation.\foot{In
Calabi--Yau compactifications of the heterotic string the spin
connection is embedded into the gauge connection. This has
received the name `standard embedding' in the traditional string
theory literature, which is more a misnomer as there is nothing
standard about this embedding. In general the spin connection is
not embedded into the gauge connection, so that H-flux is required
to satisfy the Bianchi identity. In the context of heterotic
M-theory, solutions with non-standard embedding have been
discussed in \ovrut\ and references therein.}

From the supersymmetry constraints \Strom\ (see also \Bars\ and
\Hulla) it becomes evident that the $H$-field has the geometrical
meaning of torsion of the $SU(3)$ holonomy connection. Moreover,
the $H$-field is the obstruction for the background metric to be
K\"ahler and in particular the metric turns out to be conformally
balanced\foot{See for example \michel\ for a mathematical discussion
of the balanced condition.} instead of Calabi--Yau \refs{\IPapad, \Lust,
\GauMW, \liyau, \BT}. Not being Calabi--Yau, many theorems of K\"ahler
geometry do not apply which makes their analysis more arduous. The
existence of smooth solutions has not been proven so far. It is the
purpose of this paper to fill in this gap.

In this paper, smooth flux backgrounds for the heterotic string
are constructed. The orbifold limit of these manifolds has been
described previously in the literature \refs{\DasRS, \BD, \BBDG,
\BBDGS}. The manifolds described herein are $T^2$ bundles over
a smooth four-dimensional Calabi--Yau base. We explicitly solve
the conditions imposed by supersymmetry. Moreover, consistency
requires the background to be a solution of the anomaly
cancellation condition. Of all the constraints on the background
fields, the anomaly cancellation is the most difficult one to
satisfy. It constrains the geometry as well as the gauge bundle
leading to topological restrictions. We will see that this
condition requires the base to be $K3$ and that a $T^4$ base is
prohibited for a flux compactification.

The anomaly cancellation of the heterotic theory is a highly
non-linear differential equation for the dilaton. The existence of
a smooth solution of this equation has recently been proven in
\FuYau. We will briefly describe the method used for the proof and
describe the limits placed on the dilaton field. Although our
results are derived completely within the context of the heterotic
theory, they exhibit features that should also be applicable to
flux compactifications of type II theories as string dualities map
our solution to flux backgrounds of M-theory on $K3 \times K3$ as
well as the F-theory duals discussed in \refs{\gvw, \DasRS, \gkp}.

The outline of the paper is as follows. In section 2, we set up
our notation by reviewing the supersymmetry constraints imposed on
the background in flux compactifications of heterotic strings. It
is particularly important to pay proper attention to the sign
conventions, as this is a delicate point that will have a strong
implication on the existence proof (see \Bime\ for a careful
discussion on sign conventions.). It has been pointed out by
Gauntlett et al. \GauMW\ that the ``Iwasawa solution'' presented in
\Lust\ is not a valid solution due to a sign error in the
torsional equation presented in the next section. This can easily
be seen from our derivation, as the Iwasawa solution is a $T^2$
bundle over a $T^4$ base, that will be excluded once the
Bianchi identity is taken into account. In section 3, we introduce
the conformally balanced metric ansatz and we motivate this
background using string duality which relates it to flux
compactifications of M-theory on $K3 \times K3$. We describe the
solutions for the heterotic gauge field and show that it solves
the Hermitian-Yang-Mills equation. Section 4 is devoted to showing
that the solution presented in section 3 solves the anomaly
cancellation condition. We write down the necessary topological
constraints and explain the method used in \FuYau\ to establish
existence of smooth solutions. In section 5, the properties of our
solutions as well as some concrete examples are presented. Open
problems and future directions are presented in the conclusion.

This paper is a companion paper to \FuYau\ where some of the
mathematical results described here, in particular the existence
of a smooth dilaton solution, are proven rigorously.

\newsec{Torsional constraints}

In order to set up our conventions we begin by summarizing the
supersymmetry constraints for an ${\cal N}=1$ compactification of
the heterotic string to four dimensions. The bosonic part of the
ten-dimensional supergravity action in the string frame is 
\eqn\action{ \CS = {1\o 2\k_{10}^2} \int \rd^{10}x \sqrt{-g}\,
e^{-2\phi}\left[R + 4| \partial \phi| ^2  - {1\o 2}|H |^2 - {\ap\o
4} {\rm tr} (|F|^2) \right]~, } 
where $\phi$ is the dilaton, $R$ is the curvature scalar, and
$F_{MN}$ is the gauge field strength which we take to be
hermitian, {\it i.e.} 
\eqn\ai{ F=dA - i A\w A~.} 
The three-form $H$ is defined in terms of a two-form potential $B$
and the Chern--Simons three-form $\Omega(A)$ according to
\eqn\Hdef{ H= \rd B + {\ap\o 4} \O(A) \qquad{\rm where } \qquad
\Omega(A)= {\rm tr} \left(A\w dA - i {2\o 3} A\w A\w A\right)~. }
This leads to the tree-level Bianchi identity 
\eqn\aii{\rd H =  {\ap\o 4} {\rm tr} (F\w F)~.} 
Note that beyond tree level there is an additional contribution to
$H$, namely the Lorentz Chern-Simons term, which depends on the
spin connection. This higher-derivative term is important for
anomaly cancellation \schwarz\ and will play a crucial role in our
analysis. The choice of connection to be used in the Lorentz
Chern--Simons form is a subtle issue and will be discussed in more
detail in section 4.

After including the contributions from the fermionic fields the
supergravity action is invariant under the ${\cal N}=1$
supersymmetry transformations \eqn\ai{ \eqalign{
 \d \psi_M   &  =  \nabla_M \e
+ {1\o 4}  {\bf H}_M
 \e~, \cr
  \d \l   & = \pa\!\!\!  / \phi\, \e + {1\o 2}{\bf H} \e~,\cr
 \d \c   & = 2{\bf F}  \e ~,   \cr}}
where $\psi_M$ is the gravitino, $\l$ is the dilatino and $\chi$
is the gaugino. A background is supersymmetric if a non-vanishing
spinor $\e$ satisfying $ \d_ \e ({\rm fermi})=0$ can be found.
These constraints were worked out in \Strom. We are interested in
six-dimensional Poincar\'e invariant compactifications preserving
an ${\cal N}=1$ supersymmetry in four dimensions. Since the
supersymmetry transformations {\ai} are written in the string
frame the background is a direct product of a four-dimensional
space-time and an internal six-dimensional manifold.

Unbroken supersymmetry implies the vanishing of the
four-dimensional cosmological constant and as a result the
external space is Minkowski. Moreover, the constraints imposed by
{\ai} imply the following conditions on the internal manifold $X$

\item{(1)} It is complex and the metric is hermitian. As a result
we can choose the standard local coordinates where the complex
structure $J_m{}^n$ takes the form \eqn\ax{
J_a{}^b=i\,\d_a{}^b\qquad{\rm and } \qquad J_\ba{}^\bb=-i\,
\d_\ba{}^\bb~.} The hermitian $(1,1)$-form is then related to the
hermitian metric by $J_{a\bb}=i g_{a\bb}$.

\item{(2)} It is non-K\"ahler in the presence of a non-vanishing
$H$-field, which is related to the derivative of $J$ by the
torsional constraint\foot{This corrects a sign error in \Strom.}
\eqn\Hcond{ H=i \left(\bpa-\pa\right)J~. } This condition can be
conveniently written in the form \eqn\axx{ H = d^c J~,} where we
have used the operator $d^c = i ( \bar \pa - \pa)$ which is
standard in the mathematics literature (see {\it e.g.} \grha). The
first equation in \ai\ implies that in these backgrounds spinors
can be found which are covariantly constant, not with respect to
the usual Christoffel connection, but with respect to the
`Strominger connection' which includes the $H$-flux.

\item{(3)} Moreover, there exists a holomorphic $(3,0)$-form which
we shall denote by $\O$ and which is the three-form fermion
bilinear scaled by a factor $e^{2\p}$. Its norm is proportional to
the exponential of the dilaton field, \eqn\axi{
\|\Omega\|^2=e^{-4(\phi+\phi_0)}~, } for some constant $\phi_0$.
The dilaton is, in turn, related to the metric by the condition
that $X$ is conformally balanced, {\it i.e.}
\eqn\djcon{\rd\left(\|\Omega\|\, J\w J\right)=\rd\left(e^{-2\p}\,
J\w J\right)=0~.}

\item{(4)} The gauge field satisfies the Hermitian-Yang-Mills
conditions \eqn\fcon{F^{(2,0)}=F^{(0,2)}= F_{mn}J^{mn}=0~.}

The torsional constraint on the $H$-field can be derived using the
equations of motion. The action {\action} leads to the equations
of motion \eqn\axiii{ \eqalign{ \d g^{MN}&:\quad R_{MN} +
2\,\na_M\!\na_N \phi - {1\o 4} H_{MPQ}H_N{}^{PQ}- {\ap\o 4} {\rm
tr}(F_{MQ}F_N{}^Q)=0~,\cr \d\phi& :\quad R-4|\pa \p|^2 + 4\nabla_P
\pa^P \p-{1\o 2}|H|^2 -{\ap\o 4} {\rm tr}(|F|^2)=0~,\cr \d B^{MN}&
:\quad \na_P(e^{-2\p}H^P{}_{MN})=0~,\cr \d A^M & :\quad
D_N(e^{-2\p}F^N{}_M) - {1\o 2}e^{-2\p} H_{MNP}F^{NP}=0~,\cr }}
where we have used the $\d B^{MN}$ and $\d \p$ equations to
simplify the $\d A^M$ equation of motion and Einstein equations
respectively. The trace of the Einstein equation is then
\eqn\mdvar{\nabla_M \nabla^M e^{-2\p} -
 e^{-2\p} |H|^2 - {\ap \o 4}e^{-2\p} {\rm
tr}(|F|^2)=0~.} This can be integrated over $X$ and implies
\eqn\mdchk{ \int_X e^{-2\phi} H\w \star H +{\ap \o 4} \int_X
e^{-2\phi} {\rm tr}(F\w \star F) =0~. } Note that if there are no
additional contributions, each term in \mdchk\ being positive
semi-definite must vanish identically.  But since there are $\ap
R^2$ corrections to the action that will shortly be taken into
account and that give a negative contribution to this equation, we
shall formally proceed assuming $H$ and $F$ are non-zero. Using
the fact that a Hermitian-Yang-Mills field strength satisfies
$\star F = - J\w F$ and applying \Hdef, the previous equation can
be rewritten as \eqn\mdchp{\int_X e^{-2\phi} H\w \star H - \int_X
e^{-2\phi}dH\w J =0~.} Integrating by parts we find \eqn\mddhq{
\star H= e^{2\phi}\rd(e^{-2\phi}J)~. } This is a another way of
expressing the result for $H$ which using {\djcon} can be shown to
be equivalent to \Hcond.

Together, \Hdef\ and \Hcond\ imply \eqn\jf{i\, \pa\bpa J = {\ap \o
8} {\rm tr}(F\w F)~.} Moreover, as mentioned above, beyond tree
level, the anomaly cancellation requires an additional
contribution of order $\a' R^2$ on the right hand side of \jf.
After taking this contribution into account {\jf} takes the form
\eqn\jfr{ i\, \pa\bpa J = {\ap \o 8} \left[{\rm tr} (\hR\w \hR) -
{\rm tr}(\CF\w\CF)\right]~.} Here and in the following we will be
using conventions which are standard in the mathematics
literature. Namely the curvature two-form is given by \eqn\axxiv{
\hR =\rd\hom + \hom\w\hom~, } where $\hat \omega $ is the spin
connection which will be described in more detail in section 4 and
the gauge field \eqn\axxv{ \CF =\rd \CA + \CA\w\CA~, } are both
anti-hermitian. We introduce here the calligraphic symbol ${\cal
F}$ to distinguish it from the hermitian gauge field $F$ which is
more commonly used in the physics literature. After integrating
over a four-cycle, {\jfr} requires that the first Pontryagin
numbers of the gauge and tangent bundles agree, {\it i.e.}
\eqn\axxiii{ { {p_1(E)\over 2}}={p_1(M)\over 2}~. }

\newsec{Solution ansatz and its M-theory dual origin}
In the following we present the ansatz for the metric and gauge
bundle and describe the M-theory dual of this solution.

\subsec{The metric}

We study the class of supersymmetric solutions that are
topologically $T^2$ bundles over a four-dimensional base
manifold $S$. The metric on this space can be written in the form
\eqn\bi{ ds^2 = e^{2 \phi} ds_S^2 + \left(dx+ \a_1\right)^2 +
\left( dy + \a_2\right)^2~.} Here $\phi$ depends on the coordinates
of the base manifold $S$ only, $(x,y)$ are the fiber coordinates
and $\a=\a_1 + i \a_2$ is a one-form which will be further
constrained below. Introducing complex coordinates $z=x+iy$ and
defining $\th=dz+\a$, which is required to be a $(1,0)$ form, we
can write \eqn\metric{\eqalign{ds^2= e^{2\phi} ds_{S}^2 + |dz +
\alpha |^2 ~.}} To preserve supersymmetry, we require the base
manifold $S$ to be a Calabi--Yau manifold and we denote its
K\"ahler form with $J_S$. The hermitian $(1,1)$-form on $X$ can
then be expressed through $J_S$ according to
\eqn\hemet{\eqalign{J&= e^{2\phi} J_S + \left(dx +
\alpha_1\right)\wedge \left(dy+ \alpha_2\right)\cr &= e^{2\phi}
J_S + {i\o 2}\, \th\w\bth~. }} Moreover, as we will see below, the
condition of having a conformally balanced metric requires the
two-form \eqn\thexter{ \om=\om_1+i\,\om_2 = \rd \alpha = \left(\pa
+ \bpa\right) \a = \om_S + \om_A~,} to be primitive on the base,
{\it i.e.} \eqn\bii{\om\w J_S =0~.} In the above expression,
$\om_S$ is the self-dual $(2,0)$ part of $\omega$ and $\om_A$ is
its anti-self-dual $(1,1)$ part. The holomorphic $(3,0)$-form on
$X$ is then determined to be \eqn\biv{\O=\O_S \w \th,} where
$\O_S$ is the holomorphic $(2,0)$-form on the base.

Using the previous equations, the metric can be readily checked to
satisfy the conformally balanced condition \djcon\
\eqn\confb{\eqalign{\rd(e^{-2\p}\, J\w J)&=
\rd\left(e^{-2\p}\,[e^{4\p}\,(J_S\w J_S) + i\,
e^{2\p}\,\th\w\bth\w J_S]\,\right)\cr &= i \,\rd(\th\w\bth)\w J_S
\cr &= i\,\bth\w\om\w J_S -\,i\, \th\w {\bar{\om}}\w J_S\cr
&=0~,}} where in the second line we have used that $\phi$ depends
on the base coordinates only and that $\omega$ is primitive on the
base. Note that the $e^{2\phi}$ factor in the metric precisely
cancels the $e^{-2\phi}$ factor in the conformally balanced
condition.

For an ${\cal N}=1$ compactification with non-zero $H$-flux, the
$T^2$ bundle has to be non-trivial.  A non-twisted $T^2$ fiber
would result in ${\cal N}=2$ supersymmetry in four dimensions.
Moreover, to ensure that the metric in \metric\ is
globally-defined, we impose \eqn\bv{\tom_i = {\om_i\over
2\pi\sqrt{\a'}} \in H^2(S,\IZ)~,}
%
that is $\tom_1$ and $\tom_2$ represent a non-trivial integral
cohomology class on $S$.  The normalization is due to taking the
periodicity of the torus coordinates to be \eqn\bvi{ x\sim x +
2\pi\sqrt{\a'}\qquad {\rm and } \qquad y \sim y + 2
\pi\sqrt{\a'}.} Note that $\tom=\tom_1+i\,\tom_2$ is the curvature
two-form of the $T^2$ fiber. The quantization condition is
equivalent to the requirement of the first Chern class for each
$S^1$ bundle to be integral.

The non-trivial twisting has an effect on the de Rham cohomology
of the compactification manifold $X$ \refs{\GoldP,\BBDG}. Assuming
that $\om_1$ is not proportional to $\om_2$, the second Betti
number satisfies $b_2(X)=b_2(S)-2$.  The harmonic two-forms on $X$
are those on $S$ modded out by $\om_1$ and $\om_2$, since these
two elements of $H^2(S)$ are exact in $X$. The reason is that
$\om_1=d(dx+\alpha_1)$ and similarly for $\omega_2$ . Importantly,
the area element of the $T^2$ fiber $\th\w\bth$ also does not
constitute a harmonic two-form in $X$.  By Poincar\'e duality,
this implies that the volume form of $K3$ is also not an element
in $H^4(X)$. Or equivalently, the $K3$ base is not a four-cycle of
$X$. The twisting has effectively made the volume element of $K3$
trivial in the de Rham cohomology of $X$.

\subsec{Gauge Bundle}

With the manifold being a $T^2$ bundle, we can easily construct
Hermitian-Yang-Mills bundles on the total space $X$ by taking
those on the base $S$ and pulling them back to $X$. Indeed,
suppose we have a Hermitian-Yang-Mills gauge bundle ${\cal F}^S$
on the base $S$. Then it satisfies ${\cal F}^S_{mn}J_S^{mn}=0$,
which is equivalent to \eqn\fbase{{\cal F}^S\w J_S=0~.} This in
fact implies that ${\cal F}^S_{mn}$ is also Hermitian-Yang-Mills
on $X$ since \eqn\fwhole{{\cal F}^S_{mn}J^{mn}=\star\left({\cal
F}^S\w J\w J\right) =\star\left( {\cal F}^S \w \left[
e^{4\p}\,(J_S\w J_S) + i\,e^{2\p}\,J_S\w\th\w\bth
\;\right]\right)=0~,} where \fbase\ has been used.

The obvious question is therefore whether all Hermitian-Yang-Mills
connections on $X$ are those that are lifted from the base. To
answer this, we first point out the relation between
Hermitian-Yang-Mills connections and gauge bundles which are
stable. In general, for a compact hermitian manifold $X$, a
holomorphic gauge bundle $E$ with field strength $\CF$ is called
stable if and only if all coherent subsheaves\foot{Sheaves
generalize the notion of vector bundles and allow the type of
the fiber to change (or even degenerate) over the base.  For an
accessible account, see \refs{\ESharpe}.} $E'$ of $E$ satisfy the
condition 
\eqn\stable{{\rm slope}(E') <\, {\rm slope}(E)~,}
where the slope of $E$ is defined using the degree of $E$
\eqn\degree{{\rm slope}(E)={{\rm deg~}E\o  {\rm rank}~E}= {1\o
{\rm rank}~E}\left({1\o 2\pi}\int_X {\rm tr}(\CF)\w \tJ^2\right)
~,}
with the rank being the dimension of the fiber. Here, $\tJ$ is
the Gauduchon hermitian form which in six dimensions satisfies
\Gaud \eqn\gaud{\pa\bpa \,\tJ^2=0~.} The balanced condition \djcon\
implies that the Gauduchon two-form is given by $\tJ=e^{-\phi}J$.

Now, due to a theorem of Li and Yau \LiYau, it turns out that a
vector bundle admits a Hermitian-Yang-Mills connection if and only
if it is stable (see also \Lubke). Thus, finding all
Hermitian-Yang-Mills connections on $X$ is equivalent to
categorizing the stable gauge bundles on $X$. Moreover, since we
have $\star \,(\CF_{mn}J^{mn})=\CF\w J^2=0$, we are specifically
interested in stable bundles of degree zero.\foot{To be precise, for
zero degree stable bundle, the stability requirement \stable\ should
be modified to {slope$(E')\leq\,$ slope$(E)$}.  This is known as the semistable
condition.}  As shown in section 5.3, using the anomaly cancellation
condition and also allowing 
for possible holonomy along the fibers \Verb, the relevant stable
bundles for the $T^2$ bundle over the Calabi--Yau base consist
only of the stable bundles on $S$ tensored with a holomorphic line
bundle on $X$, i.e. $\CF = \CF^S \otimes {\bf 1} + {\bf 1} \otimes
\CF^L\,$.

\subsec{M-theory dual}

The construction of a conformally balanced metric for a heterotic
flux background was first obtained via duality from M-theory
compactifications on $K3\times K3$.  The metric was first written
down in the orbifold limit in \DasRS\ and such backgrounds have
since been studied extensively in \refs{\BD,\BBDG,\BBDGS,\BT}. The
metric and the $H$-flux are derived by applying a chain of
supergravity dualities valid only at the orbifold limit of
$K3\times K3$.  The resulting geometry in the heterotic theory is
a $T^2$ bundle over the $K3$ orbifold $T^4/\ZZ_2$.  The
orbifold limit has the advantage that the form of the metric can
be written down explicitly, but has the drawback that the geometry
and the $H$-field are singular at the $16$ orbifold fixed points.
Analyses are then typically separated into consideration far from
the singularities and that at the singularities.

The class of heterotic metrics  \metric\ can be motivated via
duality from M-theory. For $S=K3$, the heterotic solution is dual
to M-theory on $K3\times K3$, with the second $K3$ taken as a
$T^4/\ZZ_2$ orbifold. To be precise, the metric is conformal to
$K3\times K3$. Starting from M-theory on $Y=K3\times K3$ with
non-zero flux, the series of dualities leading to the heterotic
solution are roughly as follows. Treat the second $K3=T^4/\ZZ_2$
as an elliptic fibration over $CP_1$. Reducing the $T^2$ fiber to
zero size, we obtain the type IIB theory on $K3 \times T^2/\ZZ_2$,
where $\ZZ_2 = \Omega(-1)^{F_L}I_{89}$ with $I_{89}: (x,y)\to
(-x,-y)$ and $\Omega$ being the world sheet parity operator.
Applying further two T-dualities, one in each direction of
$T^2/\ZZ_2$, results in the dual type I theory on $K3$ with a $T^2$
bundle. Finally, an S-duality takes the type I background
to the above heterotic solution.

The knowledge of the dual backgrounds in type IIB and M-theory is
very useful in providing insights into the heterotic flux
background. From the the dual type IIB theory on $K3 \times
T^2/\ZZ_2$, we see the origin of the twisting of the $T^2$
bundle.  Here, the $T^2$ metric is not twisted.  However, there
are non-zero $B$-fields present that under the two T-dualities are
absorbed into the metric and thus twist the $T^2$ fiber. In the
type II theory, the three-form $H=dB$ satisfies the Dirac
quantization condition. This condition leads to the requirement
that $\tilde \om_i$ are in integral cohomology classes.
Specifically, the $B$-field has the form \eqn\thfield{ B = {1\o 2}
(\a \w d\bz + {\bar \a} \w dz)\qquad {\rm where} \qquad d\a=
\omega_1 + i \omega_2~.} As in the notation of the heterotic
solution, $z$ is the complex coordinate on the $T^2$. Dirac
quantization requires that the corresponding three-form satisfies
\eqn\dirac{{1\o (2\pi)^2\alpha'}\int_{\G} \, H \in \ZZ\qquad {\rm
where} \qquad \G\in H_3\left(K3\times T^2/\ZZ_2, \ZZ\right)~.}
Notice that \dirac\ contains $\alpha'$ and thus the quantization
of $d\a$ is relative to the length-scale set by $\alpha'$.

From the dual M-theory $G$-flux background, we can obtain
insights on the heterotic anomaly cancellation equation. Indeed,
on the M-theory side the four-form $G$ is constrained by
supersymmetry and the Bianchi identity \BeckerGJ. Under duality it
maps to the heterotic three-form and Yang-Mills gauge fields. The
equation of motion associated with $G$ takes the form,
\eqn\eom{d\star G = -{1\o 2} G \w G - \beta\, X_8\qquad {\rm with
} \qquad X_8 = {1\o (2\pi)^4}{1\o 4!}\left[{1\o 8} {\rm tr} R^4 -
{1\o 32} \left({\rm tr} R^2\right)^2\right]~,} where
$\beta=2\kappa^2_{11}T_{M_2}$ is expressed in terms of the
membrane tension $T_{M_2}$ and the eleven-dimensional
gravitational constant $\kappa_{11}$. Under duality this gives
rise to the anomaly cancellation equation on the heterotic side
\dlm.

Equation \eom\ can be integrated over the compact Calabi--Yau
four-fold $Y$ to give the condition  \eqn\susycond{{1\o 2}\,
\int_Y G \w G = {\chi \o 24}~, } where $\chi$ is the Euler
character and we have $\beta =1$. Additional M2-brane sources lead
to a contributions $+N$ on the left hand side of \susycond, where
$N$ is the number of M2-branes. Since supersymmetry requires $G$
to be a primitive $(2,2)$-form, which implies self-duality, we
have \eqn\ci{ \int_Y G\w G\geq 0~, } and it vanishes only if $G=0$.
As a result a non-zero $G$-flux is consistent with a $K3\times K3$
compactification geometry in M-theory.  However, notice that the
duality mapping at each step described above does not affect the
base manifold.  Hence, if the base manifold on the heterotic side
is taken to be $T^4$, then the corresponding dual M-theory
background geometry would be $T^4\times K3$. Since $\chi(T^4\times
K3)=0$ it cannot support non-zero $G$-flux.  Thus, from the
duality perspective, there is no consistent heterotic flux
background solution with base $S=T^4$.

Finally, we describe how the heterotic gauge fields arise
from duality mapping. As discussed in \BD, the gauge fields can be
traced back to the $G$-flux in
M-theory. Their appearance can be seen most transparently in the
dual type IIB theory on $K3\times T^2/\ZZ_2$. Present at each of
the four fixed points of $T^2/\ZZ_2$ are four $D7$-branes
and one $O7$-plane. Each set of four $D7$'s supports at most a
$U(4)$ gauge bundle which is broken down to $SO(8)$ by the
projection of the $O7$-plane. These bundles are localized on
$T^2/\ZZ_2$ and hence only have dependence on the $K3$
coordinates. Applying the duality mapping, the gauge
bundles in the heterotic theory from duality at the orbifold limit have
the maximal gauge group $SO(8)^4$ and dependence only on the
base coordinates.

\newsec{Solving the anomaly cancellation}

In this section we demonstrate that the heterotic metric ansatz
\metric\ satisfies the anomaly cancellation condition
\eqn\ancan{\rd H = 2i\, \pa\bpa J = {\ap \o 4} \left[{\rm tr}
\left(\hR\w \hR\right) - {\rm tr}\left(\CF\w\CF\right)\right]. }
However, in writing this condition there is a subtlety related to
the choice of connection $\hom$ since anomalies can be cancelled
independently of the choice \Hull. Different connections
correspond to different regularization schemes in the
two-dimensional worldsheet non-linear sigma model. Hence the
background fields given for a particular choice of connection must
be related to those for a different choice by a field redefinition
\Sen. In the following we will be using the hermitian connection
\Strom. The advantage of this choice is that it implies that ${\rm
tr} (\hR\w \hR)$ is a $(2,2)$-form while the $(3,1)$ and $(1,3)$
contributions vanish. This is consistent with the other two terms in
\ancan\ which are both $(2,2)$-forms. We will use the hermitian
connection below and denote the hermitian curvature two-form
simply as $R$.

To evaluate the constraints imposed by the anomaly cancellation
condition \ancan, it is convenient to rewrite the flux and
curvature dependent terms. First we notice that the flux dependent
term can be rewritten as 
\eqn\dHasd{\eqalign{dH&= 2i\, \pa\bpa
e^{2\phi}\w J_{S} + \om_S\w \bom_S - \om_A\w\bom_A\cr &= 2i\,
\pa\bpa e^{2\phi}\w J_{S} + \om_S\w \star\bom_S + \om_A\w
\star\bom_A\cr &= 2i\, \pa\bpa e^{2\phi}\w J_{S}+
\left(\|\om_S\|^2 + \|\om_A\|^2\right){J_{S}^2 \o 2!}~,}}
where we have used the definition of $ \|\om\|$ given in the appendix. For
${\rm tr}\,R\w R$ we refer to the calculation presented in \FuYau\
which gives \foot{Note that our metric convention differs slightly from that
of \FuYau, i.e. $g_{a\bb}= (1/2)(g_{a\bb})_{FY}$.} 
\eqn\rrex{{\rm tr}\, R\w R = {\rm tr}\, R_S\w R_S +2\,
\pa\bpa \left[ e^{-2\p} {\rm tr}\,\left(\bpa \A \w \pa \A^\dagger
g^{-1}_S\right)\right] + 16 \,\pa\bpa\, \p \w \pa\bpa \, \p~ ,}
where $R_S$ and $g_S$ are respectively the hermitian curvature tensor and
the metric on $S$, and we have defined a column vector $\A$ locally
given by 
\eqn\di{\A=\left(\matrix{\A_1\cr
\A_2}\right)\qquad {\rm with } \qquad \bpa(\,\A_1 \,dz^1 + \A_2
\,dz^2)=\omega_A~ .} Here $(dz^1,dz^2)$ is the basis of
$(1,0)$-forms on $S$.

\subsec{Topological conditions}

Using the previous results we can now derive constraints on the
allowed flux background solutions. These constraints can be
obtained by integrating the anomaly cancellation equation over $X$
or the base $S$. Indeed, we can apply to \ancan\ an exterior
product with the hermitian form $J$ and integrate over the
six-manifold $X$. The three terms that contribute can be written
as follows. First, the contribution coming from the flux takes the
form 
\eqn\hjwe{\int_X 2 i\, \pa\bpa J \w J = {1\o 2}\,\int_X
e^{-4\p}(\|\om_S\|^2 + \|\om_A\|^2)\, J^3~.} 
The term involving the curvature takes the form \eqn\rrwe{\int_X {\rm
tr}\, R\w R \w  J = \int_X {\rm tr}\, R_S\w R_S \w J~,} since the
$\pa\bpa$-exact terms in \dHasd\ and \rrex\ when wedged with $J$
integrate to zero over $X$. For the gauge field term we use the six-dimensional
identity 
\eqn\Fid{\star\CF = {1\o 4}\, (J\w J) J_{mn}\CF^{mn} -
{1\o 2}J\w {\tilde \CF}, \qquad {\rm where } \qquad {\tilde
\CF}_{mn} = 2J_{mr}J_{ns} \CF^{rs}~.} 
If we now impose the supersymmetry requirement that $\CF$ is a
$(1,1)$-form, we can rewrite ${\tilde \CF}_{a\bb} = 2\, \CF_{a\bb}$
and $\star\CF = - J\w \CF$ to obtain 
\eqn\ffj{\int_X {\rm tr}\, \CF\w \CF \w J = -
\int_X {\rm tr}\, \CF \w \star \CF > 0~.} 
This expression is positive because  ${\rm tr}\,\CF \w \star \CF$ is negative
semi-definite since $\CF$ is anti-hermitian. Altogether, we obtain
the inequality 
\eqn\sixcon{\int_X {\rm tr}\, R_S\w R_S \w J =
{2\o\a'}\int_X e^{-4\p}(\|\om_S\|^2 + \|\om_A\|^2)\, J^3\, -
\,\int_X {\rm tr}\, \CF \w \star \CF~ >0~, } 
which gives a constraint for ${\rm tr}\,  R_S\w R_S$,
a four-form defined on the base manifold. Both terms on the right
hand of this equation are bigger than zero for a non-trivial
solution. As a result, backgrounds with a $T^4$ base only lead to
trivial solutions for which the fluxes, gauge fields and the twist
vanish, because for $T^4$ the curvature vanishes $R_s=0$. In
particular this implies that the Iwasawa manifold is not a
consistent heterotic flux background. Moreover, since the base is
required to be a Calabi--Yau manifold, it can only be $K3$. This
result is dual to the M-theory tadpole constraint where
non-vanishing fluxes are only allowed on manifolds with non-zero
Euler characteristic.

By integrating \ancan\ over the base manifold $K3$ we obtain the
topological constraint on which the existence proof of \FuYau\ is
based. Using \dHasd\ and \rrex\ the integrated equation takes the
form 
\eqn\ifcan{{1\o \a'} \int_S (\|\om_S\|^2 + \|\om_A\|^2) J_S\w
J_S = {1\o 2} \int_S {\rm tr}\, R_S\w R_S - {\rm tr}\, \CF\w \CF~.
} 
Multiplying both sides by a factor of $(1/4\pi^2)$ we obtain
\eqn\fourcon{\int_S \left(\|\tom_S\|^2 + \|\tom_A\|^2\right) J_S\w
J_S = -p_1(S) + p_1(E) ~>0~,~} since the integral on the left hand
side is positive definite. For a base manifold $S=K3$ the
characteristic classes satisfy \eqn\aai{2 \,c_2(K3)=-p_1(K3)=48~.}
Therefore from \fourcon\ an important equation that is at the
heart of the existence theorem derived in \FuYau\ can be obtained
\eqn\kfcon{-{p_1(E)\o 2} + \int_S \left(\|\tom_S\|^2 +
\|\tom_A\|^2\right) {J_S\w J_S \o 2!} = 24 ~.} As we will discuss
below, as long as this equation is satisfied the existence of a
smooth solution for the dilaton can be established. Note that
\eqn\abii{ p_1(E)=2~{\rm ch}_2(E) = c_1^2(E)- 2c_2(E)~,} 
and for a gauge bundle admitting spinors, $c_1(E)$ is divisible by two
\disgr. The norm of
$\tom_i$ appearing in \kfcon\ can be found from the intersection
numbers of $K3$. Since $p_1(E)<0$, the number of different allowed
gauge bundles is finite and we can write the possible solutions in
terms of the data $(\tom_1,\tom_2, E)$. Below we will explicitly
construct examples of backgrounds satisfying \kfcon.

\subsec{Differential equation and the elliptic condition}

The anomaly condition leads to the differential equation
\eqn\anomdiff{{2i\o \alpha'}\, \pa\bpa e^{2\p} \w J_S - {1\o 2}\,\pa\bpa
\left[ e^{-2\p} {\rm tr}\left(\bpa \A \w \pa \A^\dagger
g^{-1}_S\right)\right] - 4\,\pa\bpa\, \p \w \pa\bpa \, \p + \psi\,
J_S^2/2 =0~, } 
where we have defined $\psi$ according to
\eqn\psieqn{\psi\, J_S^2={1\o \alpha'}\left(\|\om_S\|^2 +
\|\om_A\|^2\right) J_S^2-  {1\o 2}\left( {\rm tr}\, R_S\w R_S -
{\rm tr}\, \CF\w \CF\right)~.} 
From the topological constraint
\ifcan, we see that $\psi$ integrates to zero on $K3$, {\it i.e.}
${\int_{K3}\,\psi = 0\,}$.  The $\psi$ term can be heuristically
treated as a source term contribution to the differential
equation. Equation \anomdiff\ is then the differential equation
that determines the functional form for the background dilaton
field $\phi$.  We will now describe the existence proof showing
that the dilaton differential equation does indeed have a
solution.

To prove that a solution to \kfcon\ indeed exists, an elliptic
condition given below is imposed. From the mathematical point
of view, this allows the application of powerful techniques for
solving elliptic non-linear partial differential equations.
However, such a condition can also be motivated from the physics
point of view. Indeed, consider the deformation of the dilaton
field $\phi\to \phi+\d\phi$ with all other background fields
fixed. For an infinitesimal variation, the deformation is studied
by linearizing \anomdiff\ with respect to $\d\phi$.  We expect the
number of independent deformations to be finite and thus it is
natural to require that the resulting second-order linear partial
differential equation for $\d \phi$ to be elliptic.\foot{A simple
example of an elliptic equation is the Laplace equation on a
torus, whose solution is a constant. As opposed to this, the wave
equation is hyperbolic and the solutions are given by an infinite
number of propagating modes.} Here ellipticity means that the
coefficient matrix of the second order derivative of $\d \phi$ is
positive. From the variation of $\d\p$, we obtain the condition
\eqn\ellip{{4\o \alpha'}e^{2\p} J_S - i\,e^{-2\p}{\rm
tr}\left(\bpa \A \w \pa \A^\dagger g^{-1}_S\right)+ 8i\, \pa\bpa\,
\p > 0~.} 
In addition, as a convention, we will choose to
normalize the volume of $K3$ to be one, {\it i.e.} $\int_{K3}
J_S^2/2=1$ and define the constant $A$ according to
\eqn\psinorm{A=\left(\int_{S}\, e^{-8\p} {J_S\w J_S \o
2!}\right)^{1/4}.} Below we will see that the solutions are
labelled by different values of $A$.

\subsec{Existence and a priori bounds }

The existence of a smooth solution for $\p$ in the differential
equation \anomdiff\ is proven in \FuYau\ using the standard
continuity method.\foot{This is the same method that established
the existence of the Calabi--Yau metric in \Yau.}
The idea is to connect via a parameter $t\in [0,1]$, a difficult
non-linear differential equation at $t=1$ to a simpler one at
$t=0$ with known solution.  Specifically for \anomdiff, consider
the one parameter family of differential equations
\eqn\diffeq{L_t(\p_t) = {2i\o \alpha'}\, \pa\bpa e^{2\p_t} \w J_S
- {t\o 2}\,\pa\bpa \left[ e^{-2\p_t} {\rm tr}\left(\bpa \A \w \pa
\A^\dagger g^{-1}_S\right)\right] - 4\, \pa\bpa \p_t \w \pa\bpa \,
\p_t + t \, \psi\, J_S^2/2 =0~.} At $t=0$, the solution is given
by the constant $\p_0=-{1\o2}\ln A\,$ which satisfies the
normalization \psinorm. The goal is to show that there also exists
a solution $\p_t$ at $t=1$ which is the differential equation \anomdiff.

To do so define the set \eqn\Tdef{T = \{t\in [0,1]\, |~
L_t(\p_t)=0\,~{\rm has~a~solution}\},} consisting of values of the
parameter $t$ for which a solution exists.  Having already a
solution for $t=0\,$, the existence of a solution at $t=1$ (that
is $t=1\in T$) is guaranteed if we can show that the set $T$ is
both open and closed. This is because the only non-empty subset
with $t\in [0,1]$ that is both open and closed is the whole set
$t= [0,1]$ which contains $t=1$.  Below we briefly describe the
standard method to show that $T$ is open and closed.

Demonstrating openness is usually not difficult. We need to show
that for any point $t_0\in T$ its neighboring points $t+\d t$ is
also in $T$.
Here, we can re-express \diffeq\ as a function of both $t$ and
$\p_t$, 
\eqn\tildeL{\tL(t,\p_t)\!=\!\star_S\!\left\{ {2i\o \alpha'}
\pa\bpa e^{2\p_t} \w J_S - {t\o 2}\,\pa\bpa \!\left[e^{-2\p_t}{\rm
tr}\left(\bpa \A \w \pa \A^\dagger g^{-1}_S\right)\right]  -
4\,\pa\bpa \p_t \w \pa\bpa \p_t + t \, \psi\, J_S^2/2\right\},}
where the Hodge $\star_S$ is with respect to the base $S=K3$.
Assuming now that $\tL(t_0,\p_{t_0})=0$ is a solution, we need to
show that the first order partial derivative $\pa \tL/\pa
\p|_{(t_0,\p_{t_0})}$ is invertible ({\it i.e.} isomorphic between
function spaces).  If so, then the implicit function theorem (see
for example \Lang) implies the existence of a connected open
neighborhood around $t_0$ that also satisfy $\tL(t,\p_t)=0$ and
hence openness.  Note that since $\pa \tL/\pa
\p|_{(t_0,\p_{t_0})}$ is a linearized differential the elliptic
condition is important for demonstrating invertibility.

The major task of the existence proof in \FuYau\ is to demonstrate
closedness by deriving the delicate estimates for $\p_t$.  Recall
that the set $T$ is closed if for any convergent sequence
$\{t_i\}$ in $T$, the limit point $t'$ is also contained in $T$.
Since the sequence $\{t_i\}$ is in $T$, there is a corresponding
sequence of functions $\{\p_{t_i}\}$ that are solutions, {\it
i.e.} $L(t_i,\p_{t_i})=0$.  Proving $T$ is closed therefore
requires that the sequence of functions $\{\p_{t_i}\}$ converges
in some Banach space to some function $\p'$ and that
$L(t',\p')=0$, {\it i.e.} $\p'=\p_{t'}$.  The sequence
$\{\p_{t_i}\}$ will converge if we can show that any solution
$\p_t$ must satisfy certain bounds that are $t$ independent.  More
explicitly, the norm (in some suitable Banach space) of $\p_t$ and
derivatives of $\p_t$ should have finite upper bounds.  These
bounds on the solutions are called a priori estimates since they
are obtained prior to and without any explicit solution.  The
bounds are characteristics of the differential equation and do
depend on $\psi$, and $A$. To show that the solution is smooth
requires only the existence of bounds up to the third derivatives
of $\p_t$.  Higher derivatives bounds can then be obtained by
applying Schauder's interior estimates (see for example Chapter~6
in \GilT). With the required boundedness, the Arzela--Ascoli
theorem (see for example \JostB) then implies that the sequence
$\{\p_{t_i}\}$ must contain a uniformally convergent subsequence.
The corresponding convergent subsequence in $\{t_i\}$ necessarily
converges to $t'$ and the limit of the subsequence $\p'$ becomes
just $\p_{t'}$. Thus, closedness is established once the difficult
estimate bounds are obtained.  We refer the reader to \FuYau\ for
details of these important estimate calculations.

\newsec{Analysis of the Solutions}

The existence proof demonstrates that the $T^2$ bundle over a
$K3$ base leads to a flux background for the heterotic string as
long as the topological condition \kfcon\ is satisfied.  Below, we
describe how the anomaly cancellation constraint restricts the
dilaton field, the twists in the $T^2$ bundle, and the stable gauge
bundles of the solution.

\subsec{Dilaton}

As worked out in \FuYau\ (see Proposition 21), a sufficient
condition for ensuring the validity of the estimates necessary to
prove the existence of a solution is \eqn\Aless{ A \ll~ 1~.} As a
result this is also the sufficient condition to demonstrate the
existence of a smooth dilaton field.\foot{More specifically, $A$
is required to be smaller than certain complicated bounds. Eq.
\Aless\ effectively ensures that the those bounds for $A$ are
satisfied. In other words, $A \ll 1$ need not be necessary but is
sufficient to guarantee a solution.} This condition corresponds to
a lower bound (see Proposition 20 in \FuYau) for \eqn\aaii{
e^{2\phi}\gg 1~ .} Note that $e^{2\phi}$ is the conformal factor
in the metric \metric\ for the $K3$ base.  A large conformal
factor implies that the volume of the base is large.  This is
consistent with duality since a large conformal factor corresponds
to a large warp factor and a large Calabi--Yau volume in the dual
M- and type IIB pictures. This warp factor is often not taken into
account since in the large volume limit it is constant to leading
order. Here, via duality, the results on the heterotic side imply
the existence of a warp factor function away from the large volume
limit in type II and M-theories.

Moreover the string coupling constant on the heterotic side
$g_{\rm s}=e^\phi$ is large. A large $g_{\rm s}$ background for
the SO(32) heterotic can be equivalently considered as a small
$g_{\rm s}$ background in the S-dual type I theory. If we consider
the $E_8 \times E_8$ heterotic instead, then the dual is M-theory
on $S^1/\IZ_2$ with the radius of the $S^1$ proportional to the
coupling $g_{\rm s}$. We note that the existence proof for $\p$
holds for both heterotic theories independently of the gauge
group.

It is worthwhile to point out that there is a one parameter family
of solutions. Indeed, the supersymmetry constraints \Hcond-\fcon\
are invariant under a constant shift of $\phi\to \phi +c$.
However, this constant shift is not an invariant of the anomaly
equation \anomdiff. Nevertheless, there is still a one parameter
family of solutions for $\phi$ labelled by $A$ in \Aless. For each
value of $A \ll 1$, there exists a smooth solution of the dilaton
for the metric ansatz \metric.  A variation of $A$ will result in
a non-constant variation of $\phi$.  But notice that for our
specific metric ansatz \metric, the supersymmetry constraints
\Hcond-\fcon\ are in fact invariant for any functional variation
of $\phi$.

\subsec{Solutions with trivial gauge fields}

The background solutions must satisfy the topological constraint
\kfcon\ which we write as \eqn\topcond{-{p_1(E)\o 2}+N = 24 \qquad
{\rm where } \qquad N=\int_{K3} (\tom_S\w \btom_S - \tom_A\w
\btom_A)~ .}
On $K3$ there is a standard basis of two-forms in the integral
cohomology class which we denote with $\tom_I$ with
$I=1,\ldots,22$. The intersection matrix is given by the integral
\eqn\interK{d_{IJ}= \int_{K3} \tom_I \w \tom_J~.} The matrix
$d_{IJ}$ is the metric of the even self-dual lattice with
Lorentzian signature $(3,19)$ given by \eqn\selfdual{ (-E_8)
\oplus (-E_8) \oplus \left(\matrix{0&1\cr 1&0}\right)\oplus
\left(\matrix{0&1\cr 1&0}\right)\oplus\left(\matrix{0&1\cr
1&0}\right)} where \eqn\Eeight{ E_8=\left( \matrix{ \hfill
2&\hfill 0&\hfill -1&\hfill 0&\hfill 0&\hfill 0&\hfill
0&\hfill0\cr \hfill0&\hfill2&\hfill0&\hfill-1 &\hfill 0&\hfill
0&\hfill 0&\hfill 0\cr \hfill -1&\hfill 0&\hfill 2&\hfill
-1&\hfill  0&\hfill 0&\hfill 0&\hfill 0\cr \hfill 0&\hfill
-1&\hfill -1&\hfill 2&\hfill -1&\hfill 0&\hfill  0&\hfill 0\cr
\hfill 0&\hfill  0&\hfill 0&\hfill -1 &\hfill 2&\hfill -1& \hfill
0& \hfill 0\cr
 \hfill 0& \hfill 0& \hfill 0& \hfill 0&\hfill -1&\hfill 2&\hfill
 -1&\hfill  0\cr
 \hfill 0& \hfill 0&\hfill  0&\hfill  0&\hfill 0&\hfill -1&\hfill
 2&\hfill -1 \cr
\hfill  0&\hfill  0& \hfill 0& \hfill 0& \hfill 0& \hfill 0&\hfill
-1&\hfill 2 }\right)} is the Cartan matrix of $E_8$ Lie algebra.
With the lattice being even, $N$ is an even positive integer and
allowed to have the maximum value of $N=24$ if the gauge bundle is
trivial.

To be more explicit, we shall write $\tom_S$ and $\tom_A$ in terms
of a basis of integral two-forms.  First, for $\tom_S$, it must be
proportional to the unique holomorphic $(2,0)$-form $\O_S$ on
$K3$. Therefore, we can write \eqn\oms{\tom_S = m\, \O_S = m\,
(\O_{S1} + \, i \, \O_{S2})~,} where $m=m_1 +\, i\,m_2$ is a
gaussian integer and we have decomposed $\O_S$ into its real and
imaginary parts.  Since $\tom_S$ is in the integral class, the
holomorphic $(2,0)$-form must be normalized as follows.
\eqn\osnorm{\int_{K3} \O_S\w \bO_S = \int_{K3} (\O_{S1}\w\bO_{S1}
+ \O_{S2}\w\bO_{S2}) = 4~.} We can similarly express
$\tom_A=\tom_{A1}+\,i\, \tom_{A2}$ and decompose
\eqn\oma{\tom_{Ai}=\sum_{I=1}^{19}\, n_i^I K_I~,} where $i=1,2$ and
$K_I$ is a basis generating the integral anti-self-dual
$(1,1)$-forms.  We note that such a basis is only present for
Kummer $K3$ surfaces \grha.  All together, we have for $N$ the
condition \eqn\tognorm{ N = 4\,(m_1^2 + m_2^2) - \sum_{IJ\,i}
d_{IJ}\, n_i^I n_i^J~, } where now the intersection matrix $d_{IJ}$
for the integral anti-self-dual forms is just
\eqn\interseA{d_{IJ}=(-E_8)\oplus (-E_8) \oplus
-2\left(\matrix{1&0&0\cr 0&1&0 \cr 0&0&1}\right)} Many solutions
can be found for the 40 integers combinations $(m_1, m_2, n_1^I,
n_2^I)$ for $N\leq 24$.  As an example, for the case of trivial
gauge bundle \eqn\aaiii{ (m_1,m_2, n_1^{19},n_2^{19})=(\pm 2, \pm
1, \pm 1, \pm 1)~,} give $N=24$.  We note that having trivial gauge
bundle requires at least one $n_i^I$ is non-zero.

\subsec{Solutions with non-trivial gauge fields}

We now consider solutions with non-trivial gauge fields.  The
gauge fields are Hermitian-Yang-Mills which as mentioned are in
one-to-one correspondence with stable bundles.  The anomaly
cancellation equation further restricts the type of bundles to
those with zero field strength in the directions of the $T^2$
fiber.  This implies that the stable vector bundles on $T^2$ bundle
over $K3$ are the stable bundles on $K3$ tensored with a 
line bundle on $X$.  The line bundle simply comes from the flat
connections (with possible twisting) on the torus fiber.  The
arguments below are similar to those given in \Verb.

We first introduce the vielbeins, $\th^1, \th^2$, and $\th^3=\th$,
which provide a local basis of orthonormal $(1,0)$-forms.  The
hermitian form is then written simply as \eqn\Jorth{J= {i\o 2}
\sum_{i=1,2} \th^i\w\bth^i \,+ \,{i\o 2}\,\th\w\bth~.} In this
basis, the $(1,1)$-form gauge field strength decomposes as
follows: \eqn\CFdecomp{\eqalign{\CF &= {i\o 2} \sum_{i=1,2} a_i\,
(\th^i\w\bth^i) + {i\o 2}\, b \,(\th^1\w\bth^2 + \th^2\w\bth^1)
\cr & \qquad\qquad  + {i\o 2}\sum_{i=1,2} b_i \,(\th^i\w\bth +
\th\w \bth^i) + {i\o 2}\,a \,(\th\w\bth) ~,}} where the
coefficients $a, a_i, b$, and $b_i$ take values in the Lie algebra
of the gauge group.  Now consider the four-form ${\rm tr}\, \CF\w
\CF$.  From the anomaly cancellation equation and the explicit
calculations of the terms $dH$ and ${\rm tr}\, R\w R$, ${\rm tr}\,
\CF\w \CF  $ can not have any dependence on $\th$ or $\bth$.
Denoting $J'_S= {i\o 2} \sum_{i=1,2} \th^i\w\bth^i\, $, we thus
have the condition \eqn\ffjs{ {\rm tr}\,(\CF\w\CF) \w J'_S = -{i\o
4} {\rm tr} \left[-b_1^2 - b_2^2 + a(a_1 + a_2)\right]
(\th^1\w\bth^1\w\th^2\w\bth^2\w\th\w\bth) =0~.} Using the
Hermitian-Yang-Mills condition, $\CF_{mn}J^{mn}=0$, which with
\Jorth\ and \CFdecomp\ imply $a_1+a_2+a=0$, we have the condition
\eqn\ffjss{{\rm tr} \left[-b_1^2 - b_2^2 -a^2\right] =0~ .} With
the gauge generators being anti-hermitian, the trace of each term
is non-negative and therefore, we have $a=b_1=b_2=0$.  Referring
back to \CFdecomp, we find that $\CF$ does not have any non-zero
components tangential to the $T^2$ fiber, {\it i.e.} $\CF_{z
m}=\CF_{\bz m}=\CF_{z\bz}=0$.

However, with  $\pi_1(T^2)= {\ZZ \times \ZZ}$ on the fiber, we can
have $U(1)$ line bundles which are non-trivial on the base coordinates.
These gauge fields can take the form $\CA=i\,p\,{(dx+\a_1)}+i\, q\,
(dy + \a_2) $ implying $\CF={i\,p\, \omega_1 + i\,q\, \omega_2}$ where
$p$ and $q$ are constants.\foot{For arbitrary constants $p$ and $q$,
these gauge fields have non-trivial holonomy along the $T^2$ bundle.  However,
imposing the topological condition \kfcon, $p$ and $q$ must then be
quantized and the holonomy along $T^2$ becomes trivial.}  Note that the
field strength does not have any components in the fiber direction.
(With the holomorphic condition, we will require that $\om_1$ and
$\om_2$ consist only of the anti-self-dual $(1,1)$-part.)  Tensoring
these $U(1)$ line bundle with the stable bundle from the $K3$ surface
gives the most general stable bundle on $X$.

Below, we give some examples of solutions that satisfy the topological
constraint \kfcon.  We will utilize degree zero stable bundles on
$K3$. A sufficient condition for the existence of a stable bundle
$E$ with $(r, c^2_1(E), c_2(E) )$ on $K3$ is given by the inequality
\refs{\Mukai,\Yosh} 
\eqn\Kbunex{2r\,c_2(E) - (r-1)\,c_1^2(E) - 2r^2 \geq -2~,} 
where $r$ is the rank of the bundle.\foot{Note
that the stable bundle with $c_1=0$ has zero degree.  If a stable
bundle with field strength $\CF$ has non-zero degree, then we can
obtain a zero degree semistable bundle by considering $\CF - {1\o r}{\rm
tr}(\CF){\bf 1}\,$.} From this condition, many possible
gauge groups are allowed.  With non-trivial gauge bundle and
twisting, the solutions can be described by the following
parameters $(r, c^2_1, c_2, m_1, m_2, n_1^I, n_2^I)$ satisfying the
topological constraint \topcond\ (inserting \abii\ and \tognorm)
\eqn\allparm{\left(c_2 - {c_1^2\o 2}\right) + 4\,(m_1^2 + m_2^2) -
\sum_{I, J, i}d_{IJ}\, n_i^I n_i^J = 24~,} 
where the intersection matrix $d_{IJ}$ is that in \interseA\ and
moreover \Kbunex\ implies 
\eqn\bconst{c_2 - {c_1^2 \o 2}~ \geq ~r - {2+c_1^2 \o
2r}~.} 
Thus for instance, consider a degree zero $SU(4)$ stable
bundle on $K3$ with $(r,c^2_1,c_2)=(4,0,4)$.  The constraint
\allparm\ can be satisfied by the twisting $(m_1,m_2)=(\pm 2, \pm
1)$).  If we consider instead $(r,c^2_1,c_2)=(4,0,20)$, then for
example we can have $(m_1,m_2)=(\pm 1, 0)$.

\newsec{Conclusion}

In this paper, we have constructed and discussed the properties of
a class of smooth compact flux backgrounds for heterotic string
theory. This is the first such solution which is tractable and is
formulated away from the orbifold limit. The existence of a smooth
dilaton solution has been proven if the topological constraint
\kfcon\ is satisfied. We have discussed in detail the properties
of the solutions, in particular those of the geometry as well as
the gauge fields. It turns out that the gauge fields do not
satisfy the `standard embedding' condition and this raises the
interesting possibility of enlarging the class of gauge symmetry
breaking patterns of heterotic strings that leads to
standard-model like models. We have presented concrete examples in
which the solutions of the Hermitian-Yang-Mills equation are given
by $SU(4)$ gauge groups but larger groups like $SU(5)$ certainly
also provide solutions. This represents a way of breaking $E_8$
down to groups like $SO(10)$ or $SU(5)$ rather than $E_6$ and
could have very interesting applications to phenomenology. We
leave the exploration of these ideas to future work.

In the following, we will discuss additional open questions and
future directions. First, it would be interesting to study
generalizations of the class of solutions presented in this paper.
So for example, we have considered a metric ansatz \bi\ which is a
torus bundle over a $K3$ base. These solutions are special
since the complex structure of the torus has been set to a
constant. A natural generalization would be to consider
non-constant $\tau$ given, for example, as the solution of
\eqn\aav{ \bar \partial \tau (z_i , \bar z_i)=0~,} where $\tau$
depends only on two of the coordinates of the base which we denote
by $z$ and $\bar z$. However, as discussed in \GreeneYA, the
solutions will necessarily be singular resulting in a
decompactified solution. Whether solutions with a non-constant
$\tau$ exist remains an open question.

Next, it would be interesting to describe supersymmetric cycles
within the torsional background geometry. These can probably be
found by representing $K3$ as an elliptic fibration over a
two-sphere. The torus fiber together with one of the circles of
the torus representing the fiber of $X$ over the base $S=K3$ is a
candidate for a supersymmetric three-cycle. Performing three
T-dualities fiberwise may give rise to a mirror symmetric
background along the lines of \StromingerIT.

Also, it would be desirable to find a precise description of the
coordinates on the moduli space for the torsional backgrounds.
However, the most interesting models are perhaps torsional
backgrounds with no moduli at all. Indeed, the existence of such
backgrounds could be motivated by using the duality map to
M-theory compactified to three dimensions. It was observed in
\AspinwallAD\ that for generic flux compactifications of M-theory
on $K3 \times K3$, all the moduli can be fixed by a combination of
fluxes and instanton effects. Studying the instanton effects on
torsional backgrounds and fixing all the moduli should be very
interesting for the construction of realistic models of particle
phenomenology with predictive power.

To conclude, it is believed that the moduli spaces of Calabi-Yau
manifolds form a connected web with the connection points given by
conifold singularities. These singularities should correspond to
points in which supersymmetric cycles collapse. Are torsional
backgrounds a part of this web? Can we describe conifold
transitions in Calabi-Yau manifolds which lead to backgrounds with
vanishing $b_2$? Can the transitions be described by the torsional
backgrounds analyzed herein? At this moment torsional backgrounds
are mainly terra incognita in the string theory landscape. The
answer to these questions may lead us to the path which connects
string theory to our four-dimensional world.

\bigskip\bigskip\bigskip
\centerline{\bf Acknowledgements}
\medskip

It is a pleasure to thank A.~Adams, K.~Dasgupta, J.~Lapan, J.~Li,
G.~Moore, T.~Pantev, R.~Reinbacher, E.~Sharpe, A.~Strominger,
A.~Tomasiello, C.~Vafa, Y.-S.~Wu, and X.-P.~Zhu for helpful discussions.
K.~Becker is supported in part by NSF grant PHY-0244722, an Alfred
Sloan Fellowship, Texas A\&M University and the Radcliffe
Institute for Advanced Study at Harvard University. M.~Becker is
supported in part by NSF grant PHY-0354401, an Alfred Sloan
Fellowship, Texas A\&M University and the Radcliffe Institute for
Advanced Study at Harvard University. K.~Becker and M.~Becker
would like to thank the Physics Department at Harvard University
and the Radcliffe Institute for Advanced Studies for hospitality.
J.-X.~Fu is supported in part by NSFC grant 10471026.
L.-S.~Tseng is supported in part by the Univerisity of Utah, the
Department of Mathematics at Harvard University and NSF grant
PHY-0244722. He would also like to acknowledge travel support from
Texas A\&M University and the kind hospitality of the High Energy
Theory Group at Harvard University. S.-T.~Yau is supported in part
by NSF grants DMS-0244462, DMS-0354737, and DMS-0306600.

\bigskip\bigskip\bigskip

\centerline{\bf Appendix}

\medskip

In this appendix we summarize our notation and conventions

\item{$\bullet$} For $p$-form tensor fields $F_{N_1 \dots N_P}$ we
define
$$
|F|^2 = {1\o p!} F_{N_1 \dots N_p} F_{M_1 \dots M_p} g^{N_1 M_1}
\dots g^{N_p M_p}~,
$$
and
$$
\eqalign{ & {\bf F}  = {1\o p!} F_{N_1 \dots N_p} \g^{N_1 \dots
N_p}~, \cr & {\bf F}_{N} ={1\o (p-1)!} F_{N N_1 \dots N_{p-1}}
\g^{N_1 \dots N_{p-1}}~, \cr & \vdots}
$$
where
$$
\g^{N_1 \dots N_p} = {1\o p!} \left(\g^{N_1} \dots \g^{N_p} \pm
{\rm permutations}\right)~.
$$

\item{$\bullet$} The gauge field $F_{MN}$ can be written in terms
of the hermitian generators $\l^a$ in the vector representation of
the $G=SO(32)$ gauge group 
\eqn\aiii{ F_{MN}=F^a_{MN}\l^a\qquad {\rm with } \qquad
a=1,\ldots,{\rm dim}(G)~.} 
This gives the generator independent result ${\rm tr}(
F_{MN}F^{MN}) = 2 F_{MN}^a F^a{}^{MN}$. Here we have used the
normalization ${\rm tr} (\l^a\l^b)=2\d^{ab}$ for generators in the
vector representation of $SO(32)$.  If $\l^a$ are in the adjoint
representation, {\rm tr} is replaced by ${1\o 30} {\rm Tr}$ since
${1\o 30}{\rm Tr} (\l^a\l^b)= {\rm tr}(\l^a\l^b)$. For the case
that the gauge group is $E_8 \times E_8$, the generators are in
the adjoint representation.

\item{$\bullet$} $R$ is the Ricci scalar constructed from the
metric $g_{MN}$ using the Christoffel connection. We are using
Lorentzian signature $(-,+,+,\ldots,+)$. We will be denoting the
curvature tensors constructed using the Christoffel connection
with $R$, $R_{MN}$, {\it etc.}

\item{$\bullet$} We have introduced the covariant derivative
$$
D_N=\na_N - i [A_N,~~]~.
$$

\item{$\bullet$} We follow the convention standard in the
mathematics literature for the Hodge star operator.  In
particular, $(\star H)_{mnp}= {1\o
3!}\,H_{rst}\,\epsilon^{rst}{}_{mnp}$ with $\epsilon_{mnprst}$
being the Levi-Civita tensor.

\item{$\bullet$} We use the definition for $\|\om \|^2$
$$
\om\,\w\, (\star_S) \bom  = \|\om\|^2 {J_S^2 \o 2!}~ .
$$

\listrefs

\end